**Did Mileva Marić assist Einstein in writing his 1905 path breaking papers?**

Galina Weinstein[*]

Writers read Einstein's letter to Marić from 1901 in which he wrote: "bringing our work on relative motion to a successful conclusion!" What came afterwards was boosted by a claim that Joffe had seen the original relativity paper manuscript, and that it was signed "Einstein-Marity" (i.e., "Marić"). This drew the attention of some writers to develop a theory according to which Mileva Marić assisted Albert Einstein in solving his physics problems, but her name was left out of the published article and only Einstein's name appears in the journal as author. Historical and primary sources *do not* support this scenario.

Einstein and Marić exchanged between them letters in which Albert told Mileva his scientific thoughts and daily deeds and ambitions.[1] Fifty-four out of the Einstein and Marić letters were found. The letters open a window to Albert and Mileva's relationship.[2] The letters reveal that science and romance were inseparable for Einstein. Einstein used to call Mileva "Dolly" and Mileva" referred to him as "Johny". Most of the love letters of Einstein to Marić comingle enthusiasm, romantic emotions and scientific thrill. Marić's letters do not give few indications of how her scientific interests were developing.

On March 27, 1901 Einstein wrote Marić about "bringing our work on relative motion to a successful conclusion!".[3] This sentence and the publication of the love letters raised conspiracy theories: Did Mileva Marić assist Einstein in writing his 1905 path breaking papers? Various people speculated about Marić's role especially in the development of the theory of relativity. Very quickly Feminists scholars entered the game. Senta Troemel-Ploetz wrote for instance, "We see in the two life stories the familiar patterns that lead to the construction of success for men and the deconstruction of success for women. It is not surprising that the editors of the *Collected Papers of Albert Einstein* have nothing more to say about Mileva Einstein-Marić than: 'Her personal and intellectual relationships with the young Einstein played an important role in his development'."[4]

The debate began when Abram Fedorovich Joffe, a member of the Soviet academy of Sciences, and later in life an assistant to Wilhelm Conrad Röntgen from 1902 until 1906, had seen the original manuscript, *Zur Elektrodynamik bewegter Korper*, and that it was signed "Einstein-Marity". Marity" is a Hungarian variant of the Serbian "Marić", Mileva's maiden name. Thus it was claimed that Mileva Marić Einstein's name was left out of the published article. Only Albert Einstein's name appears in the journal as author.

---

[*] Written while I was at The Center for Einstein Studies, Boston University

Stachel explains that Joffe's 1955 article in the Soviet journal *Uspekhi fizicheskikh nauk*, did not cite Einstein and Marić as co-authors of the 1905 relativity paper. Joffe wrote: "In 1905, three articles appeared in the 'Annalen der Physik', which began three very important branches of 20[th] century physics. Those were the theory of Brownian motion, the photon theory of light, and the theory of relativity. The author of these articles – an unknown person at that time, was a bureaucrat at the Patent Office in Bern, Einstein-Marity (Marity the maiden name of his wife, which by Swiss custom is added to the husband's family name)". Joffe did not even claim that Einstein signed his name Einstein-Marity: he just cited the Swiss custom. Daniil Semenvich Danin, a popular Russian science writer, took Joffe's above words and turned them into collaboration between Einstein and Marić.[5]

Röntgen was an experimentalist and there was no reason why a theoretical paper like the relativity paper would be given to him as a referee. The members of editors in the *Annalen*, Max Planck and Paul Drude were leading theoreticians and could certainly referee the paper. Indeed they read the paper very quickly, as they referred to Einstein's 1905 paper quite immediately.[6]

In September 1906 Röntgen requested for an offprint of the 1905 published paper, presumably because he was preparing a lecture on the equations of motion of the electron. Röntgen wrote Einstein on September 18, 1906: "Very esteemed Colleague, Permit me to turn to you with two requests! In order to complete my collection of papers on electrodynamics, I would also like to have reprints of your papers. So, my first request is that you be so kind as to send me these reprints." The second request deals with Brownian motion and a question about molecular collisions.[7]

If Röntgen read the paper in 1905 he would not need the offprint a year later.

Suppose Marić's name had been on the relativity manuscript. Who erased the name in the board of the *Annelen der Physik*? The *Annelen* had no policy against papers published by women. In addition, Einstein's three path breaking 1905 papers in question contain many authorial comments in the first person *singular*. For instance,[8]

"Zum Schlusse bemerke **ich**, daß **mir** beim Arbeiten an dem heir behandelten Probleme **mein Freund** und Kollege M. Besso treu zur Seite stand und daß **ich** demselben manche wertvolle Anregung verdanke.

Bern, Juni 1905".[9]

There are 54 Einstein- Marić letters from the period 1987-1903. It is true that only ten out of Marić's letters to Einstein from 1902 or earlier have come to light, compared with 43 of his. All of this correspondence was preserved in Einstein- Marić's papers. Evan Harris Walker claimed in his letter, "Did Einstein Espouse his Spouse's Ideas?"

3to *Physics today* from 1989, "Yet only ten letters from Mileva to Albert Einstein from this period have been found. One may wonder if there were not so carefully retained [i.e., if Einstein had not destroyed them]. I cannot help but see Mileva and Albert Einstein working as a team, hoping together to achieve the kind of husband-and wife recognition that has come to Marie and Pierre Curie".[10]

In general Einstein appears to have saved practically no early letters, while other people later tended to save his, for obvious reasons, claims Stachel. But one could not select ten of Einstein's letters to Marić that would be as devoid references to physics as are hers to him. None of Marić's letters to Einstein touches on any substantive point in physics, while his to her are chock-full of substantive comments on books and articles on physics he has read as well as on his own theoretical ideas and experimental proposals.[11]

Stachel compares the couple Marie Skoldowska and Pierre Curie to the couple Einstein-Marić, and adds the couple Paul Ehrenfest and Tatiana Afanasieva to the list. All three wives were Slavs with a higher education. All three husbands came from secular backgrounds: Einstein and Ehrenfest were Jews, raised in south German urban environments (Munich and Vienna, respectively), who had yet to establish their careers when they married. In the case of the Curies and Ehrenfests there is evidence of the importance of the woman's role in their joint work, and both wives pursued a scientific career after their husbands' deaths. Marić, of course, did not pursue a scientific career either before or after her separation from Einstein.[12]

Mileva Marić did not become a physicist or a mathematician. She failed the final examinations due to poor grades in mathematics, and the "conference of examiners" at the Polytechnic therefore decided to award the diploma to Einstein, Marcel Grossman and two other candidates in section VI, but not to Fräulein Marić.[13]

It is reasonable to assume that physics aroused emotions in Einstein during the early stage of his courtship with Marić. He felt impelled to share with Marić his research, because these discoveries filled him with so much happiness and joy. Stachel gives the following example.[14] Marić informed Einstein that she was pregnant. Soon afterwards – surely a difficult time for both of them, especially when living apart and they were still not married – he opened his reply as follows: "I have just read a wonderful paper by Lenard on the generation of cathode rays by ultraviolet rays.[15] Under the influence of this beautiful piece I am filled with such happiness and such joy that I absolutely must share some of it with you. Be happy and don't fret, darling. I won't leave you and will bring everything to a *happy conclusion*"[my emphasis].[16]

Walker and Troemel-Ploetz claimed that the words "our work" in Einstein's letter to Marić of March 27, 1901 ("[…] bring our work on relative motion to a *successful conclusion*!"[my emphasis]),[17] are an evidence for the claim that Mileva solved Albert's mathematical problems and assisted him in solving his physics problems.

Stachel wrote that, "there is *no* evidence that she *was* particularly gifted mathematically, while there *is* some evidence that she was *not*".[18]

Einstein seemed to encourage Marić by telling her that he would bring everything to successful or happy conclusion. Stachel shows that the words "our work on relative motion" have been written in the emotional context, because in no places in his other letters to Marić is there mention of "our work on relative motion", Einstein always refers to his work.[19] For instance, Einstein wrote Marić on December 19, 1901, "I spent all afternoon with Kleiner in Zurich telling him *my* ideas about the electrodynamics of moving bodies, and we talked about all sorts of other physics problems. […] He advised me to publish *my* ideas on the electromagnetic theory of light of moving bodies along with the experimental method. He found the method *I* have proposed to be the simplest and most appropriate one imaginable" [my emphasis].[20]

The crucial thing according to Stachel is the following: the letters suggest that the most important role Marić played in their intellectual relationship during these years was of "a sounding board" for Einstein's ideas. Einstein had a strong need to clarify and develop his ideas in dialogue with others, a role also played on accession by his friends Michele Besso and Conrad Habicht, after he moved to Bern.[21] Seelig quoted Einstein saying that Besso was that best sounding board in whole Europe.[22] Marić was the first of a series of "sounding boards" that Einstein needed in order to help him put the fruits of his research, carried out alone and with the aid of non-verbal symbolic systems, into a form that could be communicated to others.[23] By sounding boards he meant someone capable of understanding things that Einstein explained to them, and of asking intelligent questions that could help Einstein develop his own ideas. But people such as Besso and Marić were not capable of any such effort of their own.[24]

Last but not least, Walker raises the claim which is typical of conspiracy theories: "In February 1919 the marriage of Albert and Mileva ended in an amiable divorce. Mileva received custody of the children, child support and alimony. And in an added clause of the divorce decree, Albert Einstein agreed to pay Mileva every krona of any future Nobel Prize he might be awarded. He could keep the glory, but (in a settlement that would make an LA divorce lawyer blush) she had the *prize*." After LA lawyers blushed, Walker concluded, "I find it difficult to resist the conclusion that Mileva, justly or unjustly saw this as her reward for the part she had played in developing the theory of relativity.[25]

Marić did not want to give Einstein a divorce, and she was on a physical and mental breakdown because of the divorce. She got of the delusion that they would get together. By 1919 Germany lost the war and he did not have enough German money to give her. But everybody knew he would get the Nobel Prize sooner or later. A promise in a settlement to give Marić the money of the prize also settled the divorce.

[1] Renn, Jürgen. and Schulmann, Robert, *Albert Einstein Mileva Marić The love letters*, translated by Shawn Smith, 1992, Princeton: Princeton University Press, Introduction, p. xiv.

[2] Renn an Schulmann,1992, Introduction, p. xi-xii.

[3] Einstein to Marić, March 27, 1901, *The Collected Papers of Albert Einstein Vol. 1: The Early Years,1879–1902* (*CPAE*, Vol. 1), Stachel, John, Cassidy, David C., and Schulmann, Robert (eds.), Princeton: Princeton University Press, 1987, Doc. 94; Renn and Schulmann, 1992, letter 25.

[4] Troemel-Ploetz, Senta, *Mileva Einstein-Marić: The Woman Who Did Einstein's Mathematics*. In: Women's Studies International Forum, Volume 13, Issue 5, 1990, pp. 415-432.

[5] Stachel, John, *Einstein's Miraculous Year. Five Papers that Changed the Face of Physics*, 1998/2005, Princeton and Oxford: Princeton University Press, p. Ivi-Iv.

[6] Stachel, 1998/2005, pp. lviii-lix.

[7] Röntgen to Einstein, September 18, 1906, *The Collected Papers of Albert Einstein. Vol. 5: The Swiss Years: Correspondence, 1902–1914* (*CPAE*, Vol 5), Klein, Martin J., Kox, A.J., and Schulmann, Robert (eds.), Princeton: Princeton University Press, 1993, Doc. 40.

[8] Stachel, 1998/2005, pp. lvii.

[9] Einstein, Albert, "Zur Elektrodynamik bewegter Körper, *Annalen der Physik* 17, 1, 1905, pp. 891-921; p. 921.

[10] Walker, Evan Harris, "Did Einstein Espouse his Spouse's Ideas?, *Physics Today*, February 1989, p.10.

[11] Stachel, John, "Latter: Stachel replies to Evan Harris Walker", *Physics Today*, February 1989, pp. 11, 13.

[12] Stachel, John, "Albert Einstein and Mileva Marić: A Collaboration that Failed to Develop", 1996, in Pycior, Helena M., Slack, Nancy G. and Abir-Am G. Pnina, (eds), *Creative couples in the Sciences*, pp. 207-219; reprinted in Stachel, John, *Einstein from 'B' to 'Z'*, 2002, Washington D.C.: Birkhauser, pp 39-55; p. 48.

[13] *CPAE*, Vol 1, Doc 67.

[14] Stachel, 1996 in Stachel (2002), p. 46.

[15] Lenard, Philipp, "Erzeugung von Kathodenstrahlen durch ultra- violettes Licht", *Annalen der Physik* 2 (1900), pp. 359-375.

[16] Einstein to Marić, probably May 28, 1901, *CPAE*, Vol. 1, Doc. 111; Renn and Schulmann, 1992, letter 36.

[17] Einstein to Marić, March 27, 1901, *CPAE*, Vol. 1, Doc. 94; Renn and Schulmann, 1992, letter 25.

[18] Stachel, John, "The Young Einstein: Poetry and Truth", Talk delivered at the AAAS Session on "The Young Einstein" New Orleans, February 18, 1990, in Stachel (2002), pp. 21-38; p. 30.

[19] Stachel, 1996, in Stachel (2002), pp. 46-47.

[20] Einstein to Marić, letter 47, Dec 19 1901, *CPAE*, Vol. 1, Doc. 130, p. 328, Renn and Schulmann, 1992, letter 47.

[21] Stachel, 1996, in Stachel (2002), pp. 44-45.

[22] Seelig Carl, *Albert Einstein: A documentary biography*, Translated to English by Mervyn Savill 1956, London: Staples Press, p. 71; Seelig Carl, *Albert Einstein; eine dokumentarische Biographie*, 1954, Zürich: Europa Verlag, p. 85.

[23] Stachel, John, "Albert Einstein", *The New Dictionary of Scientific Biography*, Vol 2, Gale 2008, pp. 363-373; p. 364.

[24] Stachel, 1998/2005, p. xxxv.

[25] Walker, Evan Harris, "Did Einstein Espouse his Spouse's Ideas?, *Physics Today*, February 1989, p.11.